\documentclass[]{tCPH2e}

\begin{document}
\doi{10.1080/0010751YYxxxxxxxx}
 \issn{1366-5812}
\issnp{0010-7514}

\jvol{00} \jnum{00} \jyear{2008} \jmonth{June}

\markboth{M. Tsubota}{Quantum turbulence}


\title{{\itshape Quantum turbulence: From superfluid helium to atomic Bose-Einstein condensates}}

\author{Makoto Tsubota$^{\ast}$\thanks{$^\ast$Corresponding author. Email: tsubota@sci.osaka-cu.ac.jp
\vspace{6pt}} \\\vspace{6pt} {\em{Department of Physics, Osaka City University, Osaka 558-8585, Japan}}\\\vspace{6pt}\received{v2.1 released April 2008} }

\maketitle

\begin{abstract}
This article reviews recent developments in quantum fluid dynamics and quantum turbulence (QT) for superfluid helium and atomic Bose-Einstein condensates.  Quantum turbulence was discovered in superfluid $^4$He in the 1950s, but the field moved in a new direction starting around the mid 1990s. Quantum turbulence is comprised of quantized vortices that are definite topological defects arising from the order parameter appearing in Bose-Einstein condensation. Hence QT is expected to yield a simpler model of turbulence than does conventional turbulence. A general introduction to this issue and a brief review of the basic concepts are followed by a description of vortex lattice formation in a rotating atomic Bose-Einstein condensate, typical of quantum fluid dynamics. Then we discuss recent developments in QT of superfluid helium such as the energy spectra and dissipative mechanisms at low temperatures, QT created by vibrating structures, and the visualization of QT.  As an application of these ideas, we end with a discussion of QT in atomic Bose-Einstein condensates.\bigskip

\bigskip
\bigskip

\end{abstract}

\section{Introduction}

Nature is filled with fluid flow, from small scales to large scales. Our daily life on a human scale constantly experiences currents of air, water, and the like. The earth sustains the ocean flows and atmospheric circulation, leading to an abundance of life. In space, stars and galaxies flow, forming many large-scale structures. Most flow in nature is actually turbulent. As the velocity increases, flow generally changes from laminar to turbulent. Turbulence has been a mystery for a long time.  It has been investigated not only in basic science, including physics and mathematics research, but also in applied sciences, such as fluid engineering and aeronautics. As Feynman said, turbulence is a problem "that is common to many fields, that is very old, and that has not been solved." \cite{Feynmanlecture} This is chiefly because turbulence is a complicated dynamical phenomenon with strong nonlinearity far from an equilibrium state. 

"Turbulence" reminds one of some sketches by Leonardo da Vinci.  He observed the turbulent flow of water and drew pictures showing that turbulence has a structure comprised of vortices of different sizes (Fig. 1). Vortices may be a key to understanding turbulence.  However, vortices are not well defined for a classical fluid, and the relationship between turbulence and vortices remains unclear. 

\begin{figure}
\begin{center}
\includegraphics[height=0.18\textheight]{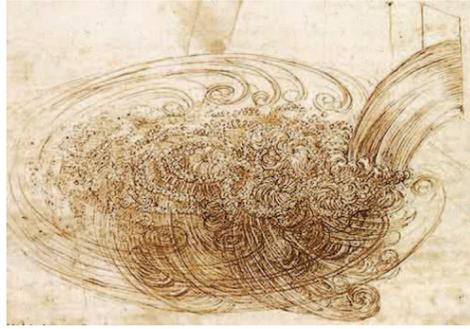}
\caption{Sketch of turbulence by da Vinci.}%
\label{Da Vinci}
\end{center}
\end{figure}

Independently of these studies in classical fluid dynamics, quantum fluids such as superfluid helium and atomic Bose-Einstein condensates (BECs) have been investigated. These systems are subject to quantum restrictions; the appearance of order parameters makes the rotational motion exist only in the presence of quantized vortices.  A quantized vortex is a stable topological defect with quantized circulation.  Such vortices give rise to quantum turbulence (QT). Since quantized vortices are well defined as elements composing a turbulent flow, QT is expected to be an easier system to study than classical turbulence (CT) and has a simpler model of turbulence.

This article addresses recent developments in quantum fluid dynamics and QT in superfluid helium and atomic BECs. In Sec. 2, basic concepts such as superfluidity, Bose-Einstein condensation, and quantized vortices are briefly reviewed. Section 3 is devoted to quantized vortices in a rotating BEC, addressing typical phenomena of quantum hydrodynamics. Section 4 describes present important topics of QT in superfluid helium. Applying these ideas, QT in atomic BECs is discussed in Sec. 5. Section 6 is devoted to conclusions.  
 
\section{Superfluidity, Bose-Einstein condensation, and quantized vortices}
Quantum fluid dynamics has been studied in connection with superfluid helium for over 50 years.
Liquid $^4$He enters a superfluid state below the $\lambda$ point (2.17 K) through
Bose--Einstein condensation of the atoms \cite{TilleyTilley}. 
The characteristics of superfluidity were discovered experimentally
in the 1930s by Kapitza {\em et al.}
The hydrodynamics of superfluid helium is well described by the two-fluid model,
in which the system consists of an inviscid superfluid (of density $\rho_s$) and 
a viscous normal fluid (of density $\rho_n$) with two independent velocity fields $\mathbf{v}_s$ and $\mathbf{v}_n$. 
The mixing ratio of the two fluids depends on temperature. 
As the system is cooled down below the $\lambda$ point, the ratio of 
the superfluid component increases, and the entire fluid becomes
superfluid below about 1 K.
The Bose-condensed system exhibits the macroscopic wavefunction
$\Psi(\mathbf{x},t)=|\Psi(\mathbf{x},t)| e^{i \theta(\mathbf{x},t)}$ as an order parameter. 
The superfluid velocity field representing the potential flow is given by $\mathbf{v}_s=(\hbar/m) \nabla \theta$ with boson mass $m$. 
Since the macroscopic wavefunction must be single-valued for space coordinate $\mathbf{x}$, the circulation $\Gamma = \oint \mathbf{v} \cdot d\mathbf{\ell}$ for an arbitrary closed loop within the fluid is quantized in terms of the quantum value $\kappa=h/m$.  A vortex with such quantized circulation is called a quantized vortex. Any rotational motion of a superfluid is only sustained 
by quantized vortices.

A quantized vortex is a topological defect characteristic of a Bose--Einstein condensate and is different from a vortex in a classical viscous fluid. First, its circulation is quantized and conserved, in contrast to a classical vortex whose circulation can have any value and is not conserved. Second, a quantized vortex is a vortex of inviscid superflow. Thus, it cannot decay by the viscous diffusion of vorticity that occurs in a classical fluid.  Third, the core of a quantized vortex is very thin, on the order of the coherence length, which is only a few angstroms in superfluid $^4$He and submicron in atomic BECs. Since the vortex core is thin and does not decay by diffusion, it is possible to identify the position of a quantized vortex in the fluid. These properties make a quantized vortex more stable and definite than a classical vortex. 

Early experimental studies of superfluid turbulence focused primarily on thermal counterflow, in which the normal fluid and superfluid flow in opposite directions. The flow is driven by an injected heat current, and it is found that the superflow becomes dissipative when the relative velocity between the two fluids exceeds a critical value \cite{GorterMellink}. Feynman proposed that it is a superfluid turbulent state consisting of a tangle of quantized vortices \cite{Feynman}. Vinen later confirmed Feynman's proposal experimentally by showing that the dissipation arises from mutual friction between vortices and the normal flow \cite{Vinen57a, Vinen57b, Vinen57c, Vinen58}. The quantization of circulation was also observed by Vinen using a vibrating wire technique \cite{Vinen61}. Subsequently, many experimental studies have examined superfluid turbulence (ST) in thermal counterflow systems, revealing a variety of physical phenomena \cite{Tough82}. The dynamics of quantized vortices are nonlinear and nonlocal, so it has not been easy to quantitatively understand the experimental results on the basis of vortex dynamics. A breakthrough was achieved by Schwarz, who clarified the picture of ST consisting of tangled vortices by a numerical simulation of the quantized vortex filament model in the thermal counterflow \cite{Schwarz85, Schwarz88}. However, since the thermal counterflow has no analogy in conventional fluid dynamics, these experimental and numerical studies are not helpful in clarifying the relationship between ST and classical turbulence (CT). Superfluid turbulence is often called quantum turbulence (QT), which emphasizes the fact that it is comprised of quantized vortices.

Comparing QT and CT reveals definite differences. Turbulence in a classical viscous fluid appears to be comprised of vortices, as pointed out by da Vinci. However, these vortices are unstable, repeatedly appearing and disappearing. Moreover, the circulation is not conserved and is not identical for each vortex. Quantum turbulence consists of a tangle of quantized vortices that have the same conserved circulation.  Looking back at the history of science, {\it reductionism}, which tries to understand the nature of complex things by reducing them to the interactions of their parts, has played an extremely important role. The success of solid state physics owes much to  {\it reductionism}. In contrast, conventional fluid physics is not reducible to elements, and thus does not enjoy the benefits of {\it reductionism}. However, quantum turbulence is different, being reduced to quantized vortices. Thus {\it reductionism} is applicable to quantum turbulence. Consequently, QT should lead to a simpler model of turbulence than CT. 

Based on these considerations, research into quantum fluid dynamics has opened up new directions
since the mid 1990s. 
One new direction has occurred in the field of low temperature physics by studying superfluid helium. It started with the attempt to understand the relationship between QT and CT \cite{Vinen02, Vinen06}.  
The energy spectrum of fully developed classical turbulence is known to obey the Kolmogorov law in the inertial range.  Recent experimental and numerical studies support a Kolmogorov spectrum in QT. Following these studies, QT research on superfluid helium has moved to important topics such as the dissipation process at very low temperatures, QT created by vibrating structures, and visualization of QT \cite{PLTP, Tsubota08}.
Another new direction is the realization of Bose-Einstein condensation in trapped atomic gases in 1995, which has stimulated intense experimental and theoretical activity \cite{Pethick02}.  As proof of the existence of superfluidity, quantized vortices have been created and observed in atomic BECs, and numerous efforts have been devoted to a number of fascinating problems \cite{KasamatsuPLTP}.  Atomic BECs have several advantages over superfluid helium.  The most important is that modern optical techniques enable one to directly control condensates and visualize quantized vortices. Because a series of experiments on BECs clearly show the properties of quantum fluid dynamics \cite{Abo01, Madison00, Madison01}, we shall start with this topic.

\section{Quantized vortices in a rotating BEC}
What happens if we rotate a cylindrical vessel with a classical viscous fluid inside? Even if the fluid is initially at rest, it starts to rotate and eventually reaches a steady rotation with the same rotational speed as the vessel. In that case, one can say that the system contains a vortex that mimics solid-body rotation.\footnote{Thin boundary layers of the Ekman and Stewartson types can appear in a rotating classical fluid, but we do not consider such complicated structures here.} A rotation of arbitrary angular velocity can be sustained by a single vortex.  However, this does not occur in a quantum fluid.  Because of quantization of circulation, superfluids respond to rotation, not with a single vortex, but with a lattice of quantized vortices. Feynman noted that in uniform rotation with angular velocity $\Omega$ the curl of the superfluid velocity is the circulation per unit area, and since the curl is $2\Omega$, a lattice of quantized vortices with number density $n_0=\text{curl} v_s/\kappa=2\Omega/\kappa$  ("Feynman's rule") arranges itself parallel to the rotation axis \cite{Feynman}.   Such experiments were performed for superfluid $^4$He: Packard {\it et al.} visualized vortex lattices on the rotational axis by trapping electrons along the cores \cite{Williams74,Yarmchuck82}.  

This idea has also been applied to atomic BECs. Several groups have observed vortex lattices in rotating BECs \cite{Abo01, Madison00, Madison01,Matthews99}. Among them, Madison {\it et al}. directly observed nonlinear processes such as vortex nucleation and lattice formation in a rotating $^{87}$Rb BEC \cite{Madison01}. 
By sudden application of a rotation along the trapping potential, an initially axisymmetric condensate undergoes a collective quadrupole oscillation to an elliptically deformed condensate. This oscillation continues for a few hundred milliseconds with gradually decreasing amplitude. Then the axial symmetry of the condensate is recovered and vortices enter the condensate through its surface, eventually settling into a lattice configuration.

This observation has been reproduced by a simulation of the Gross-Pitaevskii (GP) equation for the macroscopic wavefunction $\Psi(\mathbf{x},t)=|\Psi(\mathbf{x},t)| e^{i \theta(\mathbf{x},t)}$ in two-dimensional \cite{Tsubota02,Kasamatsu03} and three-dimensional \cite{Kasamatsu05a} spaces.
The corresponding GP equation in a frame rotating with frequency ${\bf \Omega}=\Omega \hat{\bf z}$ is given by
\begin{equation}
(i-\gamma) \hbar \frac{\partial \Psi({\bf x},t)}{\partial t}= \left[ - \frac{\hbar^{2}}{2m}\nabla^{2} +V_{\rm ex}({\bf x}) + g |\Psi({\bf x},t)|^{2} - \Omega L_{z} \right] \Psi({\bf x},t) . \label{GPeqrot}
\end{equation}
Here $V_{\rm ex}$ is a trapping potential, and $L_{z}=- i \hbar (x \partial_{y} - y \partial_{x})$ is the angular momentum along the rotational axis. The interparticle potential $V_{\rm int}$ is approximated by a short-range interaction $V_{\rm int} \simeq g \delta ({\bf x-x}')$, where $g = 4\pi \hbar^{2} a / m $ is a coupling constant characterized by the s-wave scattering length $a$.
The term $\gamma$ indicates phenomenological dissipation.\footnote{We used $\gamma=0.03$ in this simulation \cite{Tsubota02,Kasamatsu03}. } 

Figure 2 shows the profile of the condensate density  $|\Psi(\mathbf{x},t)|^2$ and the phase $\theta(\mathbf{x},t)$ when there is a quantized vortex in a trapped BEC.  The density has a hole representing the vortex core. The phase has a branch cut between 0 and $2\pi$, and the edge of the branch cut corresponds to the vortex core around which the phase rotates by $2\pi$ as the superflow circulates. One can therefore clearly identify the vortex both in the density and the phase.

\begin{figure}
\begin{center}
\begin{minipage}{100mm}
\subfigure[]{
\resizebox*{5cm}{!}{\includegraphics{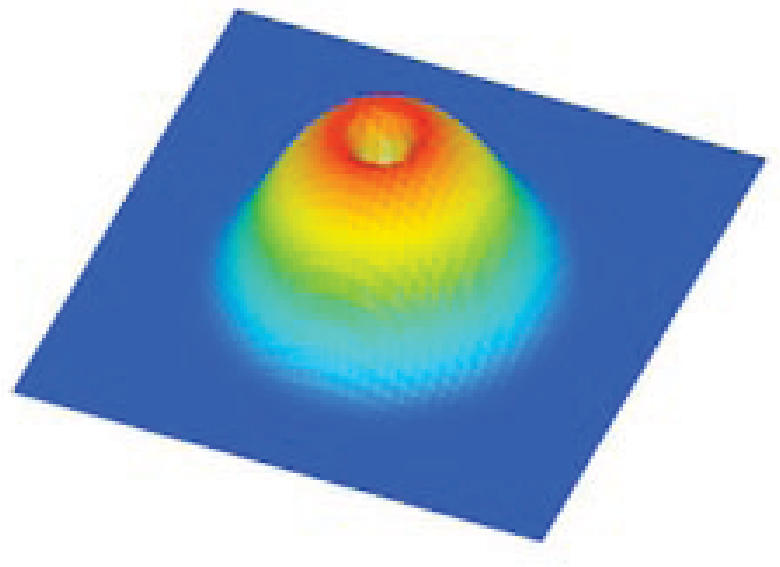}}}%
\subfigure[]{
\resizebox*{5cm}{!}{\includegraphics{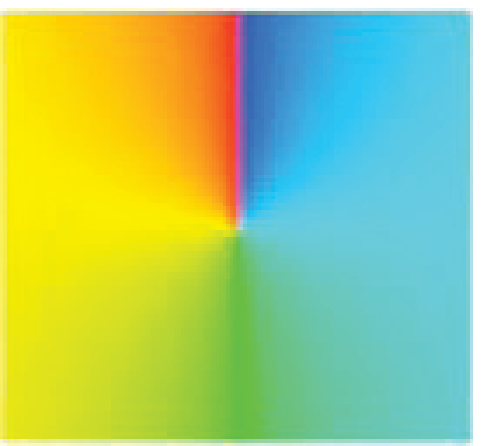}}}%
\caption{Profile of (a) the condensate density and (b) the phase of the macroscopic wavefunction when there is a quantized vortex in a trapped BEC. The value of the phase varies continuously from 0 (red) to $2\pi$ (blue).}%
\label{one vortex}
\end{minipage}
\end{center}
\end{figure}

A typical two-dimensional numerical simulation of Eq. (\ref{GPeqrot}) \cite{Tsubota02,Kasamatsu03} for the vortex lattice formation is shown in  Fig. 3, where the condensate density and the phase are displayed together.  The trapping potential is
\begin{equation} V_{\rm ex}=\frac12 m\omega^2  [ (1+\epsilon_x)x^2+(1+\epsilon_y )y^2 ], \end{equation}
where $\omega=2\pi \times 219$ Hz, and the parameters $\epsilon_x=0.03$ and $\epsilon_y=0.09$ describe small deviations from axisymmetry corresponding to experiments \cite{Madison00, Madison01}.
We first prepare an equilibrium condensate trapped in a stationary potential; the size of the condensate cloud is determined by the Thomas-Fermi radius $R_{\rm TF}$.   When we apply a rotation with $\Omega=0.7\omega$, the condensate becomes elliptic and performs a quadrupole oscillation [Fig. 3(a)]. Then, the boundary surface of the condensate becomes unstable and generates ripples that propagate along the surface [Fig. 3(b)]. As stated previously, it is possible to identify quantized vortices in the phase profile also. As soon as the rotation starts, many vortices appear in the low-density region outside of the condensate [Fig. 3(a)]. Since quantized vortices are excitations, their nucleation increases the energy of the system. Because of the low density in the outskirts of the condensate, however, their nucleation contributes little to the energy and angular momentum.\footnote{Actually the vortex-antivortex pairs are nucleated in the low-density region. Then the vortices parallel to the rotation are dragged into the Thomas-Fermi surface, while the antivortices are repelled to the outskirts.} Since these vortices outside of the condensate are not observed in the density profile, they are called  "ghost vortices".  Their movement toward the Thomas-Fermi surface excites ripples [Fig. 3(b)]. It is not easy for these ghost vortices to enter the condensate, because that would increase both the energy and angular momentum. Only some vortices enter the condensate cloud to become "real vortices" wearing the usual density profile of quantized vortices [Fig. 3(d)], eventually forming a vortex lattice [Fig. 3 (e) and (f)]. The number of vortices forming a lattice is given by "Feynman's rule" $n_0=2\Omega/\kappa$. The numerical results agree quantitatively with these observations. 

Note the essence of the dynamics. The initial state has no vortices in the absence of rotation. The final state is a vortex lattice corresponding to rotational frequency $\Omega$.  In order to go from the initial to the final state, the system makes use of as many excitations as possible, such as vortices, quadruple oscillation, and surface waves. We refer the readers to Ref. 26 for details.  

These experimental and theoretical results demonstrate typical behavior of quantum fluid dynamics in atomic BECs.

\begin{figure}[h]
\includegraphics[height=0.25\textheight]{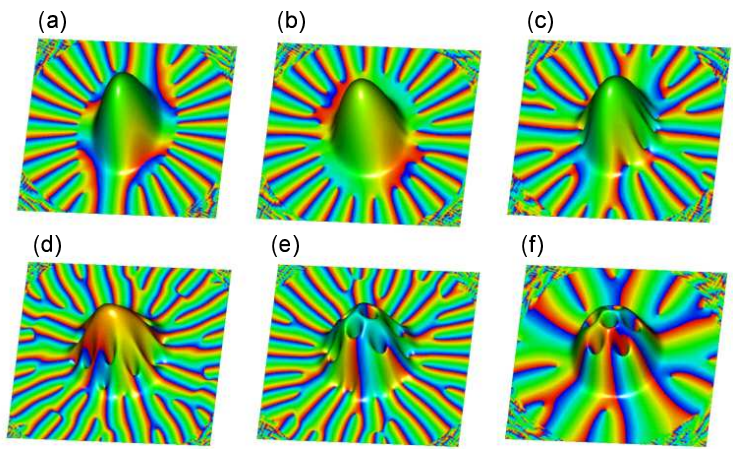}
\caption{Dynamics of vortex lattice formation in a rotating BEC. The figure simultaneously shows both the condensate density and phase. The graphs are at (a) $t=$63 ms, (b) 305 ms, (c) 350 ms, (d) 410 ms, (e) 450 ms, and (f) 850 ms after the start of the rotation. The phase varies continuously from 0 (red) to $2\pi$ (blue).}
\label{BEC}
\end{figure}

\section{Quantum turbulence in superfluid helium}
This section reviews research on QT since the mid 1990s. Before considering QT, we begin with classical fluid dynamics and statistical properties of CT \cite{Frisch}.

\subsection{Classical turbulence}
Classical viscous fluid dynamics is described by the Navier--Stokes equation,
\begin{equation}
\frac{\partial}{\partial t} \mathbf{v}(\mathbf{x},t) + \mathbf{v}(\mathbf{x},t) \cdot \nabla \mathbf{v}(\mathbf{x},t) = - \frac{1}{\rho} \nabla P(\mathbf{x},t) + \nu \,\nabla^2 \mathbf{v}(\mathbf{x},t),
\label{eq-Navier-Stokes}
\end{equation}
where $\mathbf{v}(\mathbf{x},t)$ is the velocity of the fluid, $P(\mathbf{x},t)$ is the pressure, $\rho$ is the density of the fluid, and $\nu$ is the kinematic viscosity. The flow can be characterized by the ratio of the second term on the left-hand side of Eq. (\ref{eq-Navier-Stokes}), referred to as the inertial term, and the second term on the right-hand side, called the viscous term. This ratio defines the Reynolds number $R=\bar{v}D/\nu$, where $\bar{v}$ and $D$ are a characteristic velocity and length scale, respectively. When $\bar{v}$ increases and the Reynolds number exceeds a critical value, the system changes from a laminar to a turbulent state, in which the flow is complicated and contains eddies.

Such turbulent flow shows characteristic statistical behavior \cite{Kolmogorov41a, Kolmogorov41b}. 
We assume a steady state of fully developed turbulence of an incompressible classical fluid. 
Energy is injected into the fluid at a rate $\varepsilon$ in an energy-containing range. In an inertial range, this energy is transferred to smaller length scales without dissipation. In this range, the system is locally homogeneous and isotropic, leading to energy spectral statistics described by the Kolmogorov  law,
\begin{equation} E(k)=C\,\varepsilon^{2/3}\,k^{-5/3}.\label{eq-Kolmogorov} \end{equation}
The energy spectrum $E(k)$ is defined by $E=\int d \mathbf{k} E(k)$, where $E$ is the kinetic energy per unit mass and $k$ is the wavenumber from the Fourier transform of the velocity field. The spectrum of Eq. (\ref{eq-Kolmogorov}) is derived by assuming that $E(k)$ is locally determined only by the energy flux $\varepsilon$ and by $k$. The energy transferred to smaller scales is dissipated at the Kolmogorov wavenumber $k_K=(\epsilon/\nu^3)^{1/4}$ in an energy-dissipative range via the viscosity of the fluid at a dissipation rate $\varepsilon$ in Eq. (\ref{eq-Kolmogorov}), which is equal to the energy flux $\Pi$ in the inertial range. The Kolmogorov constant $C$ is a dimensionless parameter of order unity. The Kolmogorov spectrum has been confirmed experimentally and numerically in turbulence at high Reynolds numbers. 
The inertial range is thought to be sustained by a self-similar Richardson cascade in which large eddies break up into smaller eddies through vortex reconnections. In CT, however, the Richardson cascade is not completely understood because it is impossible to definitely identify each eddy. 

\subsection{Dynamics of quantized vortices}
As described in Sec. 2, most experimental studies on superfluid turbulence have examined thermal counterflow. However, the nonlinear and nonlocal dynamics of vortices delayed progress in the microscopic understanding of the vortex tangle. Schwarz overcame these difficulties \cite{Schwarz85, Schwarz88} by developing a direct numerical simulation of vortex dynamics connected with dynamical scaling analysis, enabling the calculation of physical quantities such as the vortex line density, anisotropic parameters, and mutual friction force. The observable quantities calculated by Schwarz agree with typical experimental results.  

Two formulations are generally available for studying the dynamics of quantized vortices.  One is the vortex filament model and the other is the Gross--Pitaevskii (GP) model. 

\subsubsection{ Vortex filament model}
As discussed in Sec. 1, a quantized vortex has quantized circulation. The vortex core is extremely thin, usually much smaller than other characteristic length scales of the vortex motion. These properties allow a quantized vortex to be represented as a vortex filament. In classical fluid dynamics, the vortex filament model is an idealization. However, the vortex filament model is accurate and realistic for a quantized vortex in superfluid helium.

The vortex filament model represents a quantized vortex as a filament passing through the fluid, having a definite direction corresponding to its vorticity. Except for the thin core region, the superflow velocity field has a classically well-defined meaning and can be described by ideal fluid dynamics \cite{Schwarz85, Tsubota00}. The velocity at point $\mathbf{r}$ due to a filament is given by the Biot--Savart expression,
\begin{equation}
\mathbf{v}_{s} (\mathbf{r} )=\frac{\kappa}{4\pi}\int_{\cal L} \frac{(\mathbf{s}_1 - \mathbf{r}) \times
d\mathbf{s}_1}{|\mathbf{s}_1-\mathbf{r}|^3},
\label{BS}
\end{equation}
where $\kappa$ is the quantum of circulation.  The vortex moves with the superfluid velocity. At finite temperatures, mutual friction occurs between the vortex core and the normal flow. 

A numerical simulation method on this model has been described in detail elsewhere \cite{Schwarz85, Schwarz88,Tsubota00}. A vortex filament is represented by a single string of points separated by distance $\Delta\xi$. The vortex configuration at a given time determines the velocity field in the fluid, thus causing the vortex filaments to move. 
Vortex reconnection needs to be included when simulating vortex dynamics.  A numerical study of a classical fluid shows that close interaction of two vortices leads to their reconnection, chiefly because of the viscous diffusion of vorticity. Schwarz assumed that two vortex filaments reconnect when they come within a critical distance of each other and showed that statistical quantities such as the vortex line density are not sensitive to how the reconnections occur \cite{Schwarz85, Schwarz88}.  Even after Schwarz's study, it remained unclear as to whether quantized vortices can actually reconnect. However, Koplik and Levine directly solved the GP equation to show that two closely quantized vortices reconnect even in an inviscid superfluid \cite{Koplik93}. More recent simulations have shown that reconnections are accompanied by the emission of sound waves having wavelengths on the order of the healing length \cite{Leadbeater01, Ogawa02}. 

Schwarz numerically studied how remnant vortices develop into a statistical steady vortex tangle under thermal counterflow \cite{Schwarz88}.  The tangle is self-sustained by the competition between the excitation due to the applied flow and the dissipation by mutual friction. The numerical results were quantitatively consistent with experimental results, confirming Feynman's original picture that QT is a tangle of quantized vortices. 

The following two quantities are characteristic of a vortex tangle. The line density $L$ is defined as the total length of vortex cores per unit volume. The mean spacing $\ell$ between vortices is given by $\ell=L^{-1/2}$. 

\subsubsection{The Gross-Pitaevskii (GP) model}
In a weakly interacting Bose system, the macroscopic wavefunction $\Psi(\mathbf{x},t)$ appears as the order parameter of the Bose--Einstein condensate, obeying the Gross--Pitaevskii (GP) equation \cite{Pethick02},
\begin{equation}
i \hbar \frac{\partial \,\Psi(\mathbf{x},t)}{\partial \,t} = \biggl( - \frac{\hbar ^2}{2m}\nabla^2
+g |\Psi(\mathbf{x},t)|^{2}-\mu \biggr) \Psi(\mathbf{x},t).
\label{gpeq}
\end{equation}
Writing $\Psi = | \Psi | \exp (i \theta)$, the squared amplitude $|\Psi|^2$ is the condensate density and the gradient of the phase $\theta$ gives the superfluid velocity  $\mathbf{v}_s = (\hbar/m) \nabla \theta$, corresponding to frictionless flow of the condensate. This relation causes quantized vortices to appear with quantized circulation. The only characteristic scale of the GP model is the coherence length defined by $\xi=\hbar/(\sqrt{2mg}\,| \Psi |)$, which determines the vortex core size. 
The GP model can explain not only the vortex dynamics but also phenomena related to vortex cores, such as reconnection and nucleation. However, the GP equation is not applicable to superfluid $^4$He, which is not a weakly interacting Bose system. But it is applicable to Bose--Einstein condensation of a dilute atomic Bose gas \cite{Pethick02}, as discussed in Sec. 3.

\subsection{Modern research trends}
Most older experimental studies on QT were devoted to thermal counterflow. Since this flow has no classical analog, these studies did not significantly contribute to the understanding of the relation between CT and QT. Since the mid 1990s, important experimental studies have been published on QT in the absence of thermal counterflow, differing significantly from previous studies \cite{Vinen02,Vinen06,Tsubota08}.

The first important contribution was made by Maurer and Tabeling \cite{Maurer98}, who confirmed experimentally the Kolmogorov spectrum in superfluid $^4$He.
A turbulent flow was produced in a cylinder by driving two counter-rotating disks.
The authors measured the local pressure fluctuations to obtain the energy spectrum.  
The experiments were made at three different temperatures: 2.3 K, 2.08 K, and 1.4 K.
Both above and below the $\lambda$ point, the Kolmogorov spectrum was confirmed.

Next was a series of experiments on grid turbulence performed for superfluid $^4$He above 1 K by the Oregon 
group \cite{Stalp99, Skrbek00a,Skrbek00b,Stalp02}.  
Flow through a grid is usually used for generating turbulence in classical fluid dynamics \cite{Frisch}.
At sufficient distance behind the grid, the flow displays homogeneous isotropic turbulence.
This method has also been applied to superfluid helium.  
In the Oregon experiments, the helium was contained in a channel
along which a grid was pulled at constant velocity. 
A pair of second-sound transducers was implanted in the walls of the channel to observe vortex tangles.
In combining their observations with the decay of the turbulence, the authors had to make several assumptions.
The analysis is too complicated to be described in detail here.  
The important point is that the coupling between the superfluid and the normal fluid by mutual friction makes a quasiclassical flow appear at length scales much larger than $\ell$ and causes the fluid to behave like a one-component fluid \cite{Vinen00}. The line density is found to decay as $t^{-3/2}$.  A simple analysis shows that the  $t^{-3/2}$ decay can be derived from a quasiclassical model with a Kolmogorov spectrum \cite{Stalp99,Vinen02}.

Subsequently many experimental, theoretical, and numerical works appeared. The present areas of study of QT are \cite{PLTP, Tsubota08}: the energy spectra and the dissipation mechanism at zero temperature; QT created by vibrating structures; and visualization of QT.

\subsubsection{Energy spectra and dissipation at zero temperature}
The pioneering experiments by Maurer {\em et al.} and by the Oregon group were performed at relatively high temperatures where the normal fluid component was significant, and so the situation is complicated because of the two-fluid behavior. Here we confine ourselves to the case of pure superfluid at zero temperature. 
What happens to QT at zero temperature is not trivial \cite{PLTP}.  The first problem is the nature of the energy spectrum of the turbulence for the pure superfluid component. The second problem is the dissipation in this system. Since there is no normal fluid component, any dissipative mechanism does not work at large scales. However, some dissipative mechanism should operate at small length scales. One possibility is acoustic emission during vortex reconnections. In classical fluid dynamics, vortex reconnections cause acoustic emission.  In quantum fluids, numerical simulations of the GP model show acoustic emission at every reconnection event \cite{Leadbeater01}. However, this mechanism is thought not to be important in superfluid helium because of its short coherence length.\footnote{Acoustic emission may be more efficient in atomic BECs in which the coherence length is generally not much shorter than the system size.} Another possible mechanism is the radiation of sound (phonons) by the oscillatory motion of the vortex cores.  The third problem is how energy is transferred from large to small scales at which the dissipative mechanism operates. If the energy spectrum obeys the Kolmogorov law, the energy should be transferred through a Richardson cascade of quantized vortices in contrast to the classical case. Vortex reconnections sustaining a Richardson cascade should be less effective at size scales shorter than the mean vortex spacing $\ell$.

No experimental studies have directly addressed the energy spectra of QT at zero temperature, although three numerical studies have been made. The first study was done by Nore {\em et al.} using the GP model \cite{Nore97a, Nore97b}. They numerically solved the GP equation starting from Taylor--Green vortices and followed their time development. The quantized vortices became tangled and the energy spectra of the incompressible kinetic energy appeared to obey the Kolmogorov law for a short period, though eventually deviated from it.  The second study was done by the vortex filament model \cite{Araki02}.  Araki {\em et al.}  made a vortex tangle arising from Taylor--Green vortices and obtained an energy spectrum consistent with the Kolmogorov law. The third used the modified GP model by Kobayashi and Tsubota \cite{Kobayashi05a, Kobayashi05b}.   

The Kolmogorov spectra were confirmed for both decaying \cite{Kobayashi05a} and steady \cite{Kobayashi05b} QT by the modified GP model.  The normalized GP equation is
\begin{equation} i\frac{\partial}{\partial t}\Phi(\mathbf{x},t)=[-\nabla^2-\mu+g|\Phi(\mathbf{x},t)|^2]\Phi(\mathbf{x},t),\label{eq-GP} \end{equation}
which determines the dynamics of the macroscopic wavefunction $\Phi(\mathbf{x},t)=f(\mathbf{x},t)\exp[ i \phi(\mathbf{x},t)]$. 
The hydrodynamics in the GP model are compressible. The total number of condensate particles is $N=\int d\mathbf{x} |\Phi(\mathbf{x},t)|^2$ and the total energy is
\begin{equation} E(t)=\frac{1}{N} \int d\mathbf{x}\:\Phi^\ast(\mathbf{x},t)\Big[-\nabla^2+\frac{g}{2}f(\mathbf{x},t)^2\Big]\Phi(\mathbf{x},t),\end{equation}
as represented by the sum of the interaction energy $E_{int}(t)$, the quantum energy $E_{q}(t)$, and the kinetic energy $E_{kin}(t)$ \cite{Nore97a, Nore97b},
\begin{eqnarray} E_{int}(t)=\frac{g}{2N}\int d\mathbf{x}\:f(\mathbf{x},t)^4,  E_{q}(t)=\frac{1}{N}\int d\mathbf{x}\:[\nabla f(\mathbf{x},t)]^2, E_{kin}(t)=\frac{1}{N}\int d\mathbf{x}\:[f(\mathbf{x},t)\nabla\phi(\mathbf{x},t)]^2.\label{eq-energy-definition} \end{eqnarray}
The kinetic energy is further divided into a compressible part $E_{kin}^{c}(t)$ due to compressible excitations and an incompressible part $E_{kin}^{i}(t)$ due to vortices. The Kolmogorov spectrum is expected for $E_{kin}^{i}(t)$.

The failure to obtain a Kolmogorov law in the pure GP model \cite{Nore97a, Nore97b} is attributable to the following. The simulations showed that $E_{kin}^{i}(t)$ decreases and $E_{kin}^{c}(t)$ increases while the total energy $E(t)$ is conserved because many compressible excitations are created through vortex reconnections \cite{Leadbeater01, Ogawa02} and disturb the Richardson cascade of quantized vortices. Kobayashi {\em et al.} overcame these difficulties and obtained a Kolmogorov spectrum in QT that revealed an energy cascade \cite{Kobayashi05a, Kobayashi05b}.  They made numerical calculations of the Fourier-transformed GP equation with  dissipation,
\begin{eqnarray} (i-\tilde{\gamma}(\mathbf{k}))\frac{\partial}{\partial t}\tilde{\Phi}(\mathbf{k},t)=[k^2-\mu(t)]\tilde{\Phi}(\mathbf{k},t) +\frac{g}{V^2}\sum_{\mathbf{k}_1,\mathbf{k}_2}\tilde{\Phi}(\mathbf{k}_1,t)\tilde{\Phi}^\ast(\mathbf{k}_2,t) \times\tilde{\Phi}(\mathbf{k}-\mathbf{k}_1+\mathbf{k}_2,t). \label{eq-Fourier-GP1} \end{eqnarray}    
Here $\tilde{\Phi}(\mathbf{k},t)$ is the spatial Fourier component of $\Phi(\mathbf{x},t)$ and $V$ is the system volume. The healing length is $\xi=1/|\Phi|\sqrt{g}$.  The dissipation has the form $\tilde{\gamma}(\mathbf{k})=\gamma_0\theta(k-2\pi/\xi)$ with step function $\theta$, which only dissipates  excitations smaller than $\xi$. This form of dissipation can be justified by a coupled analysis of the GP equation and the Bogoliubov--de Gennes equations for thermal excitations \cite{Kobayashi06a}. 

Kobayashi {\em et al.} confirmed the Kolmogorov spectra for decaying turbulence \cite{Kobayashi05a}.  To obtain a turbulent state, they started the calculation from an initial configuration in which the density was uniform and the phase of the wavefunction had a random spatial distribution.  The initial state was dynamically unstable and soon developed turbulence with many vortex loops. The spectrum $E_{kin}^{i}(k,t)$ was then found to obey the Kolmogorov law. 

\begin{figure}[h] \centering \begin{minipage}{0.4\linewidth} \begin{center} \includegraphics[width=.8\linewidth]{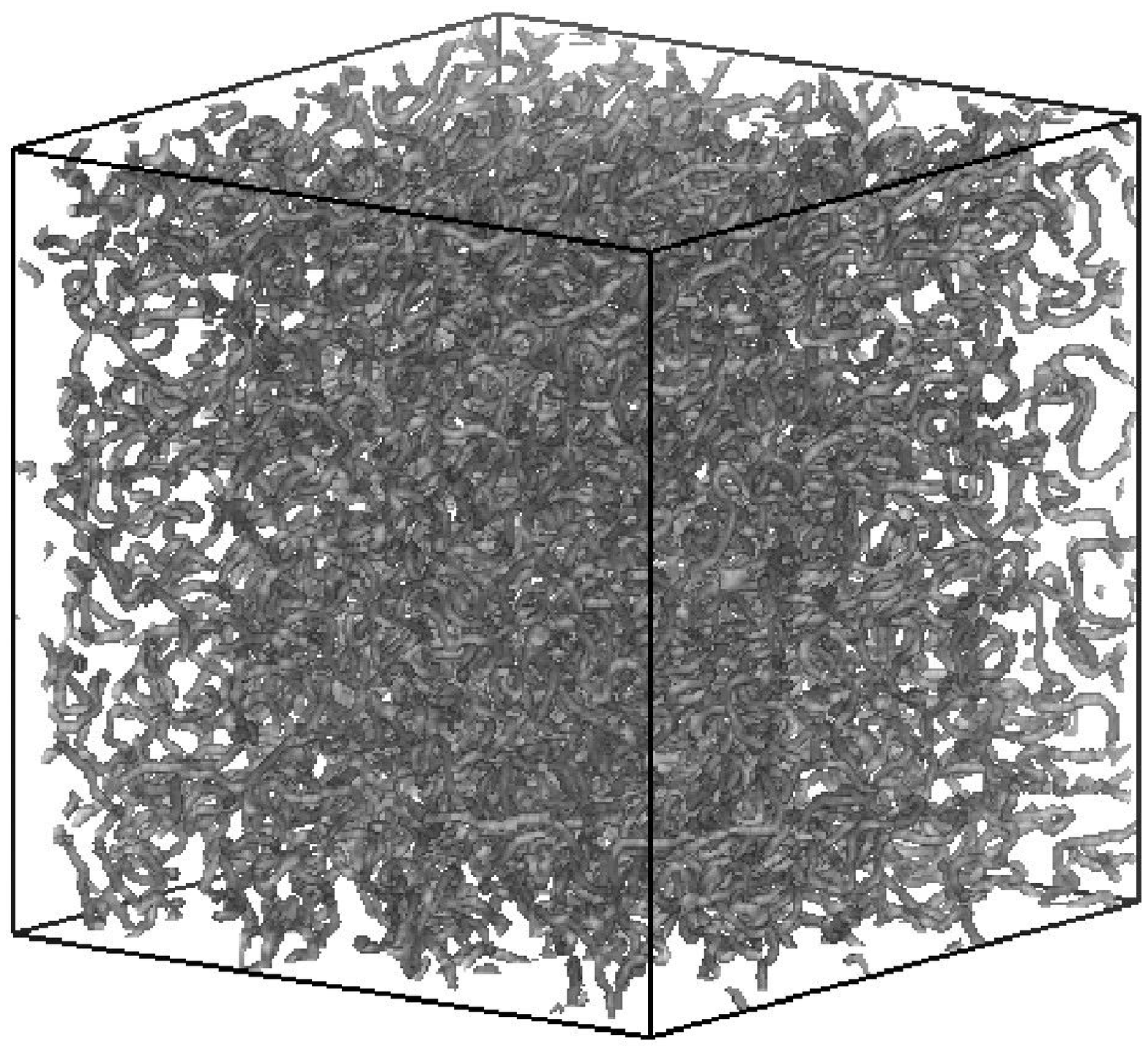}\\ (a) \end{center} \end{minipage} \begin{minipage}{0.4\linewidth} \begin{center} \includegraphics[width=.75\linewidth]{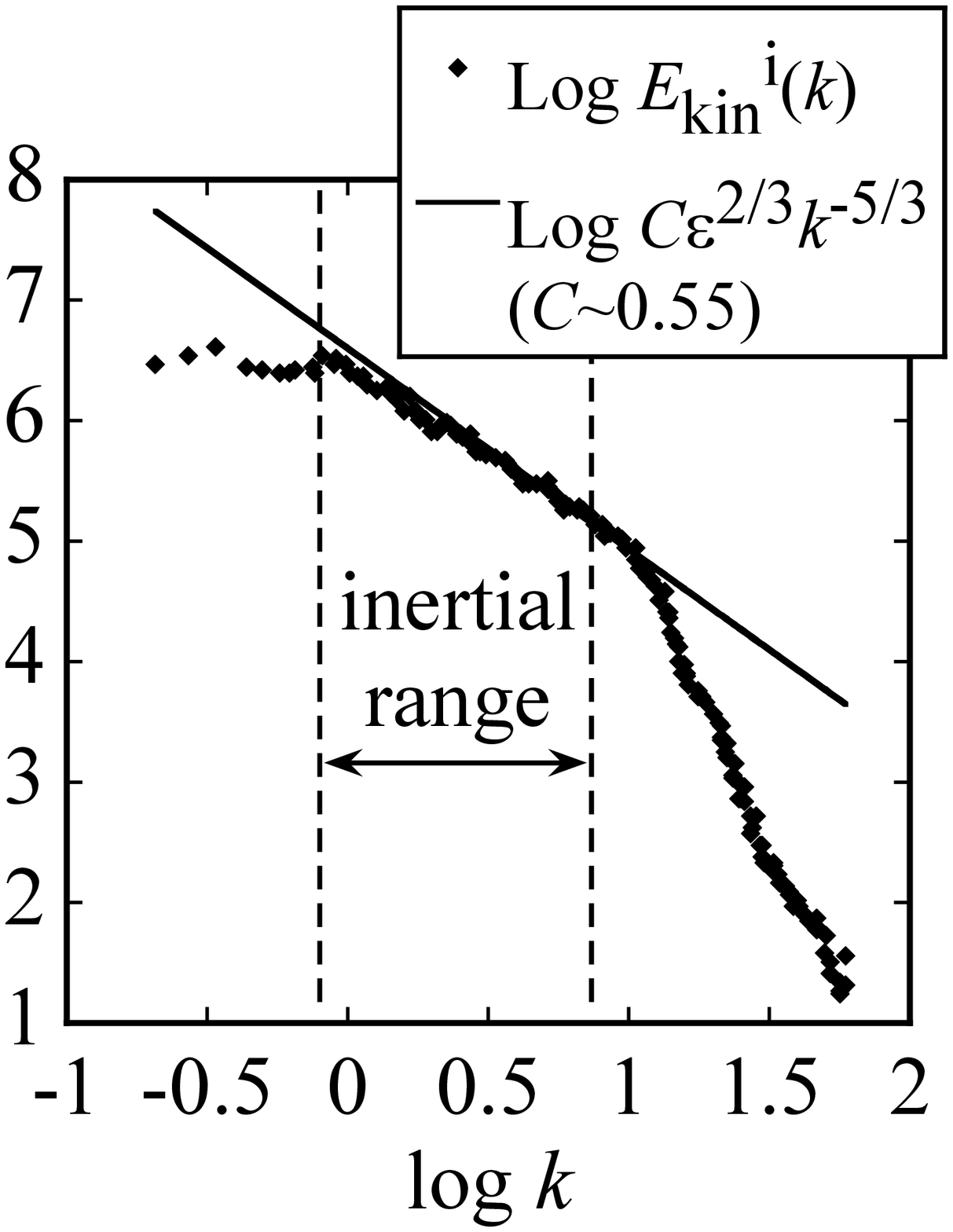}\\ (b) \end{center} \end{minipage}
\caption{(a) A typical vortex tangle. (b) The energy spectrum $E_{kin}^{i}(k,t)$ for QT. The plotted points are from an ensemble average of 50 randomly selected steady states. The solid line is the Kolmogorov law \cite{Kobayashi05b}.} \label{fig-steady-Kolmogorov} \end{figure}

A more elaborate analysis was made for steady QT by introducing energy injection at large length scales in addition to energy dissipation at small scales \cite{Kobayashi05b}. Energy injection at large scales was effected by moving a random potential $V(\mathbf{x},t)$ that had a characteristic spatial scale $X_0$.  Once steady QT occurs as a balance between energy injection and dissipation, it has an energy-containing range $k<2\pi /X_0$, an inertial range $2\pi/X_0<k<2\pi/\xi$, and an energy-dissipative range $2\pi/\xi <k$. A typical simulation of steady turbulence was performed for $X_0=4$. The dynamics started from the uniform wavefunction.  The moving random potential nucleates sound waves as well as vortices, but the incompressible kinetic energy $E_{kin}^{i}(t)$ due to vortices becomes dominant in the total kinetic energy $E_{kin}(t)$.  The energies soon become almost constant, and steady QT is obtained. 
The energy cascade can be confirmed quantitatively by checking that the energy dissipation rate $\varepsilon$ of $E_{kin}^{i}(t)$ is comparable to the energy flux $\Pi(k,t)$ via a Richardson cascade in the inertial range.  The energy flux is found to be approximately independent of $k$ and comparable to $\varepsilon$.  As shown in Fig. \ref{fig-steady-Kolmogorov} (b), the energy spectrum is quantitatively consistent with the Kolmogorov law in the inertial range $2\pi/X_0<k<2\pi/\xi$, equivalent to $0.79 < k < 6.3$.

The above arguments are chiefly limited to length scales larger than the mean spacing $\ell$ of a vortex tangle, in which the Richardson cascade is effective. What happens at smaller scales where the Richardson cascade is less effective? The most probable scenario is the Kelvin-wave cascade.  
A Kelvin wave is a deformation of a vortex line into a helix, with the deformation propagating as a wave along the vortex \cite{Thomson80}. Kelvin waves were first observed by making torsional oscillations in uniformly rotating superfluid $^4$He \cite{Hall58, Hall60}. The approximate dispersion relation for a rectilinear vortex is 
$\omega_k = (\kappa k^2)/(4\pi) (\ln(1/ka_0)+c)$ 
with a constant $c \sim 1$ and a cutoff parameter $a_0$ equal to the vortex core size. Here $k$ is the wavenumber of an excited Kelvin wave, which is different than the wavenumber used for the energy spectrum above.  At a finite temperature, a significant fraction of the normal fluid contains damped Kelvin waves because of mutual friction.  At very low temperatures, however, mutual friction does not occur and the only possible mechanism of dissipation is the radiation of phonons \cite{Vinen01}. Phonon radiation becomes effective only when the frequency is very high, typically of the order of GHz ($k \sim  10^{-1}$ nm$^{-1}$). Therefore, a mechanism is required to transfer energy to such high wavenumbers for Kelvin-wave damping.  An early numerical simulation based on the vortex filament model showed that Kelvin waves are unstable to the buildup of sidebands \cite{Samuels90}. Consequently nonlinear interactions between different Kelvin wavenumbers can transfer energy from small to large wavenumbers, which constitutes a Kelvin-wave cascade.  This idea was first suggested by Svistunov \cite{Svistunov95} and later developed  and confirmed through theoretical and numerical work by Kivotides {\em et al.} \cite{Kivotides01}, Vinen {\em et al.} \cite{Vinen03},  and Kozik and Svistunov \cite{Kozik04, Kozik05a, Kozik05b}. 

Consider the nature of the transition between the Richardson and the Kelvin-wave cascades. Turbulence at scales larger than the mean vortex spacing obeys the Kolmogorov law with a Richardson cascade. In this regime, the nature of individual vortices is not relevant, and the large-scale velocity field created collectively by the vortices mimics classical turbulent behavior. This regime may therefore be called classical. On the other hand, at length scales smaller than the mean vortex spacing the energy is transferred along each vortex through a Kelvin-wave cascade; this regime has no classical analog and can be called quantum.  Several theoretical considerations of the classical-quantum crossover have been proposed \cite{L'vov07, Kozik08a, Kozik08b}, but this topic is not yet settled and remains controversial.

\subsubsection{Quantum turbulence created by vibrating structures}
Recently, vibrating structures including discs, spheres, grids, and wires, have been used for research into QT \cite{VinenPLTP}. Despite differences between the structures, the experiments show surprisingly similar behavior. 
When a classical viscous fluid steadily flows past a structure, the flow changes from laminar to turbulence in the wake as the Reynolds number increases. The net drag force on the structure is often expressed in terms of the drag coefficient $C_D$ as $F_D=1/2 C_D\rho A v^2$, where $A$ is the projected area of the structure onto a plane normal to the flow. For laminar viscous flow, the drag is approximately proportional to $v$ so that $C_D \sim v^{-1}$, while turbulent flow with high Reynolds number yields a drag coefficient $C_D$ of order unity.  This drastic reduction in the drag force is called the drag crisis, and is closely related to the formation of the boundary layer \cite{LandauFluid}. A boundary layer in a high velocity gradient produces eddies. However, such boundary layers do not appear in the superfluid component. Another classical aspect is the effect of oscillations.  The oscillatory case is more complicated than that of steady flow because a second length scale appears in addition to the linear size of the structure, namely the viscous penetration depth $\delta = \sqrt{2\nu/\omega}$ where $\omega$ is the oscillation frequency of the structure \cite{VinenPLTP}. Owing to the importance of the topic in marine engineering, a number of experimental studies have been published on oscillatory classical flow past a cylinder. As laminar flow becomes unstable, the drag coefficient $C_D$ reduces to a value of order unity. However, the values of $C_D$ oscillate within the range 0.5 to 2.0, showing a more complicated transition to turbulence than in the case of steady flow. Visual observations show that vortices are formed in the wake of the oscillating cylinder. A few experimental studies have reported classical flow around an oscillating sphere.  A typical experiment shows that a single vortex ring is generated and shed during each half period of oscillation \cite{DonnellyPC}.

Typical behavior appeared in the pioneering observation of QT for an oscillating microsphere in superfluid $^4$He by J\"ager {\em et al.} \cite{Jager95}. The sphere they used had a radius of approximately 100 $\mu$m. It was made from a strongly ferromagnetic material and was magnetically levitated in superfluid $^4$He. Its response to an alternating drive was observed.  At low drives, the velocity response $v$ was proportional to the drive $F_D$, taking the "laminar" form $F_D=\lambda(T) v$ with a temperature-dependent coefficient $\lambda(T)$.  At high drives, the response changed to the "turbulent" form $F_D=\gamma(T) (v^2-v_0^2)$ above a critical velocity $v_0$. At low temperatures, the transition from laminar to turbulent response was accompanied by hysteresis.
Subsequently, several groups have experimentally investigated the transition to turbulence in superfluid $^4$He and $^3$He-B by using grids, wires, and tuning forks \cite{VinenPLTP}.

These experimental studies reported some common behaviors independent of the details of the structures, including its type, shape, and surface roughness. The observed critical velocities were in the range from 1 to 200 mm/s.  Since the velocity was lower than the Landau critical velocity of approximately 50 m/s, the transition to turbulence should come not from intrinsic nucleation of vortices but from extension or amplification of remnant vortices.\footnote{Superfluid $^4$He at rest usually contains remnant vortices \cite{Awschalom84}. They are metastably pinned between the walls of the container and are established when the liquid is cooled down through the $\lambda$ point to the superfluid phase.} Such behavior is seen in the numerical simulation using the vortex filament model \cite{Hanninen07}.  Figure 5 shows how the remnant vortices initially attached to a sphere develop into turbulence under an oscillating flow. Many unresolved problems remain, such as the nature of the critical velocity and the origin of the hysteresis in the transition between the laminar and turbulent response.
\begin{figure}[htb] \centering \begin{minipage}[t]{0.7\linewidth} \begin{center} \includegraphics[width=.7\linewidth]{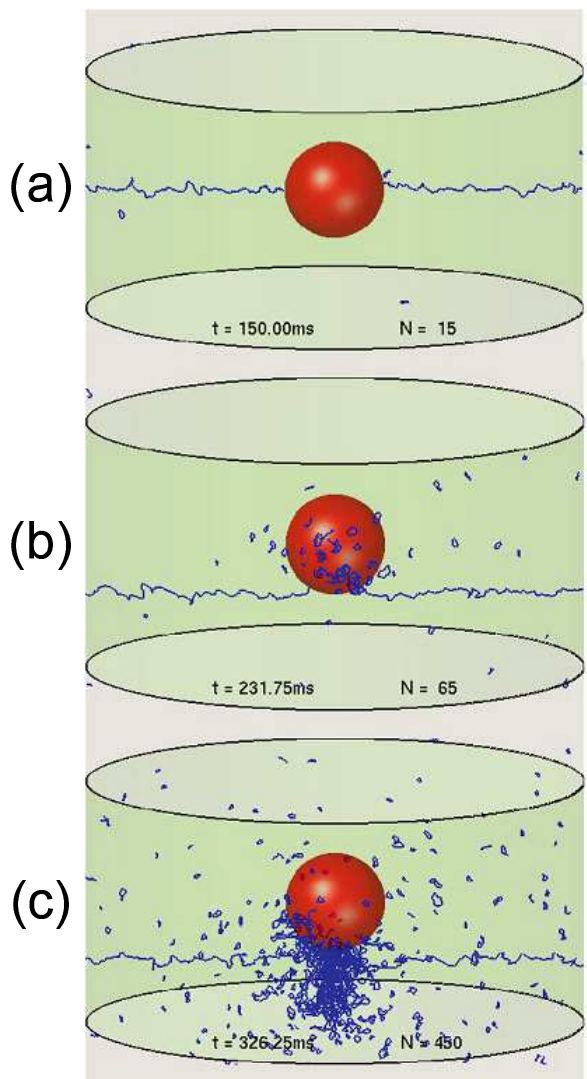} \end{center} \end{minipage} 
\caption{Evolution of a vortex line near a sphere of radius
100~$\mu$m in an oscillating superflow of 150 mm/s at 200~Hz.
[H\"anninen, Tsubota, and Vinen, Phys. Rev. B \textbf{75} (2007) 064502, reproduced with permission. Copyright 2007 by the American Physical Society.]} \label{Hanninen}\end{figure}

\subsubsection{Visualization of quantum turbulence}
There has been little direct experimental information about flow in superfluid $^4$He because standard flow visualization techniques are not applicable to it. However, the situation is changing \cite{SciverPLTP}. For QT, one can seed the fluid with tracer particles in order to visualize the flow field. Quantized vortices are observable by appropriate optical techniques.  

A significant contribution was made by Zhang and Van Sciver \cite{Zhang05}. Using the particle image velocimetry (PIV) technique with 1.7-$\mu$m-diameter polymer particles, they visualized a large-scale turbulent flow both in front of and behind a cylinder in counterflowing superfluid $^4$He at finite temperatures. In classical fluids, such turbulent structures are seen downstream of objects such as cylinders, and the structures periodically detach to form a vortex sheet. In the case of $^4$He counterflow, however, the locations of the large-scale turbulent structures were relatively stable; they did not detach and move downstream, although local fluctuations in the turbulence were evident. 

Do the tracer particles follow the normal flow, the superflow, or a more complex flow? Barenghi {\em et al.} studied this question theoretically and numerically to show that the situation changes depending on the size and mass of the tracer particles \cite{Poole05, Kivotides07, Kivotides08}.

Another important contribution was the visualization of quantized vortices by Bewley {\em et al.} \cite{Bewley06}.  In their experiments, liquid helium was seeded with solid hydrogen particles smaller than 2.7 $\mu$m in diameter at a temperature slightly above $T_\lambda$, following which the fluid was cooled to below $T_\lambda$.  When the temperature is above $T_\lambda$, a homogeneous cloud of the particles was seen that disperses throughout the fluid. However, on passing through $T_\lambda$, the particles coalesced into web-like structures.  Bewley {\em et al.} suggested that these structures represent decorated quantized vortex lines. They reported that the vortex lines form connections rather than remaining separated, and were homogeneously distributed throughout the fluid. The observed fork-like structures may indicate that several vortices are attached to the same particle. Applying this technique to thermal counter flow, Paoletti {\em et al.} have directly visualized the two-fluid behavior \cite{Paoletti08}. 

\section{Quantum turbulence in atomic BECs}
The long history of research into superfluid helium has uncovered two main cooperative phenomena of quantized vortices: vortex lattices under rotation and vortex tangles in QT. However, almost all studies of quantized vortices in atomic BECs have been limited to vortex lattices. This section briefly discusses QT in atomic BECs.

Kobayashi and Tsubota proposed an easy, powerful method to make a steady QT in a trapped BEC, by using precession \cite{Kobayashi07}.  The dynamics of the wavefunction are described by the GP equation with dissipation. First, one traps a BEC cloud in a weakly elliptical harmonic potential,
\begin{equation} U(\mathbf{r})=\frac{m\omega^2}{2}[(1-\delta_1)(1-\delta_2)x^2 +(1+\delta_1)(1-\delta_2)y^2+(1+\delta_2)z^2], \end{equation}
where the parameters $\delta_1$ and $\delta_2$ exhibit elliptical deformation in the $xy$- and $zx$-planes.  Second, to transform the BEC into a turbulent state, a rotation is applied along the $z$-axis followed by a rotation along the $x$-axis, as shown in Fig. 6(a). The rotation vector is 
$\Omega(t)=(\Omega_x,\Omega_z\sin\Omega_xt,\Omega_z\cos\Omega_xt)$
 where $\Omega_z$ and $\Omega_x$ are the frequencies of the first and second rotations, respectively. Consider the case where the spinning and precessing rotational axes are perpendicular to each other. Then the two rotations do not commute and cannot be represented by their sum.  This form of precession is also used for turbulence in water \cite{Goto07}.
 
 Starting from a stationary solution without rotation or elliptical deformation, Kobayashi {\it et al.} numerically calculated the time development of the GP equation by turning on the rotation $\Omega_x=\Omega_z=0.6$ and elliptical deformation $\delta_1=\delta_2=0.025$. By monitoring the kinetic energy and anisotropic parameters, the system eventually becomes statistically steady and isotropic. In the steady state, the spectrum $E_{kin}^{i}(k,t)$ of the incompressible kinetic energy per unit mass is consistent with the Kolmogorov law [Fig. 6(b)]. The inertial range which sustains the Kolmogorov law is determined by the Thomas-Fermi radius $R_{\rm TF}$ and the coherence length $\xi$.  The application of a combined precession around three axes enables one to obtain more isotropic QT \cite{Kobayashi08}. The velocity field in a BEC cloud can be observed by Bragg spectroscopy \cite{Stenger99}, enabling one to obtain the energy spectrum.

\begin{figure}[h]
\includegraphics[height=0.3\textheight]{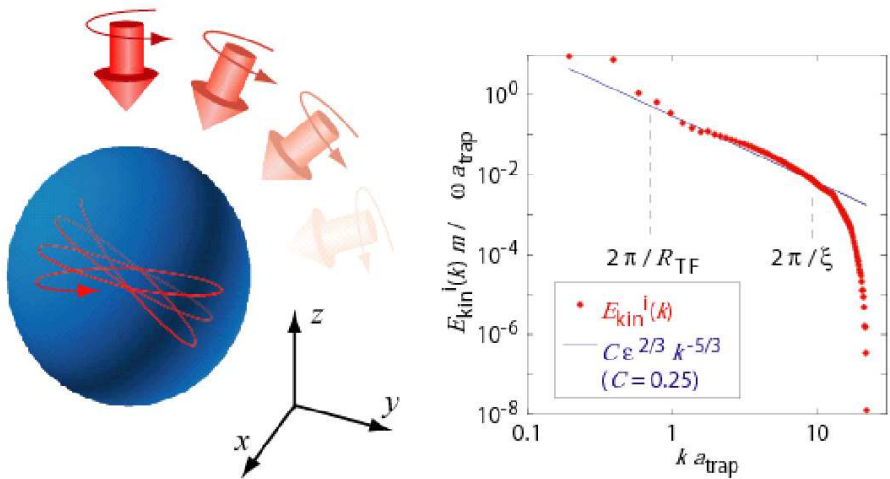}
\caption{QT in atomic BECs. (a) The BEC cloud is subject to precessions. (b) Energy spectrum of a steady QT. The dots are a numerically obtained spectrum of the incompressible kinetic energy, while the solid line is the Kolmogorov spectrum. Here, $R_{\rm TF}$ is the Thomas-Fermi radius amd $a_{\rm trap}=\sqrt{\hbar/m\omega}$ is the characteristic length scale of the trap.}
\end{figure}

There are several advantages of studying QT in atomic BECs compared to in superfluid helium. First, one can observe the vortex configuration, probably even the Richardson cascade process of vortices. In fact one can directly study the relation between the real-space Richardson cascade and the wavenumber-space cascade (the Kolmogorov spectrum). Second, one can control the transition to turbulence by changing the rotational frequencies or other parameters. For example, rotation along one axis forms a vortex lattice. If one applies another rotation, it may rotate the lattice if the frequency is low.  If the frequency is high, however, the second rotation can melt the lattice into a vortex tangle. That would enable one to investigate in detail the entire transition to turbulence. Third, by changing the shape of the trapping potential, one can study the effect of the anisotropy on turbulence; a typical question is how the Kolmogorov spectrum is changed when the BEC becomes anisotropic. This issue is related to the transition between 2D and 3D turbulence.   Fourth, one can create rich QT in multicomponent BECs by controlling the interaction parameters. Multicomponent BECs have been extensively studied and found to create many kinds of vortex structures depending on the intercomponent interaction \cite{Kasamatsu05b}. By using such systems, we will be able to make multicomponent QT coupled through the interaction. 

\section{Conclusions}
In this article, we have reviewed recent research on quantum hydrodynamics and QT in superfluid helium and atomic BECs. Research on QT is currently a key branch in low-temperature physics, attracting the attention of many scientists. It is comprised of quantized vortices as definite elements, in contrast to conventional turbulence. Thus, the investigation of QT may lead to a breakthrough in understanding a mystery of nature pondered since the era of da Vinci. There are many related topics not addressed in this article, regarding which we refer the readers to other review articles \cite{PLTP, Tsubota08}.  

\section*{Acknowledgements}

The author thanks Kenichi Kasamatsu and Michikazu Kobayashi for help with the manuscript. The author acknowledges a Grant-in-Aid for Scientific Research from JSPS (Grant No. 18340109) and a Grant-in-Aid for Scientific Research on Priority Areas from MEXT (Grant No. 17071008).

\medskip

\label{lastpage}


\begin{thebibliography}{10}
\markboth{Taylor \& Francis and I.T. Consultant}{Contemporary Physics}
\bibitem[1]{Feynmanlecture} R. P. Feynman, R. B. Leighton, and M. Sands, {\em The Feynman Lectures on Physics}, Addison-Wesley, San Francisco, 2006.
\bibitem[2]{TilleyTilley}
D. R. Tilley and J. Tilley, {\itshape Superfluidity and Superconductivity}, 3rd ed., Institute of Physics Publishing, Bristol and Philadelphia, 1990.
\bibitem[3]{GorterMellink} C. J. Gorter and J. H. Mellink, {\em On the irreversible processes in liquid helium II},  Physica 15  (1949), pp. 285--304.
\bibitem[4]{Feynman} R. P. Feynman, {\em Application of quantum mechanics to liquid helium}, in {\it Prog. Low Temp. Phys.}, ed. C. J. Gorter, North-Holland, Amsterdam, 1955, Vol. I, pp. 17--53.
\bibitem[5]{Vinen57a} W. F. Vinen, {\em Mutual friction in a heat current in liquid helium II: I. Experiments on steady heat currents}, Proc. Roy. Soc. A 240 (1957), pp. 114--127.
\bibitem[6]{Vinen57b} W. F. Vinen, {\em Mutual friction in a heat current in liquid helium II: II. Experiments on transient effects}, Proc. Roy. Soc. A 240 (1957), pp. 128--143.
\bibitem[7]{Vinen57c} W. F. Vinen, {\em Mutual friction in a heat current in liquid helium II: III. Theory of the mutual friction}, Proc. Roy. Soc. A 242 (1957), pp. 493--515.
\bibitem[8]{Vinen58} W. F. Vinen, {\em Mutual friction in a heat current in liquid helium II: IV. Critical heat currents in wide channels}, Proc. Roy. Soc. A 243 (1958), pp. 400--413.
\bibitem[9]{Vinen61} W. F. Vinen, {\em The detection of a single quantum of circulation in liquid helium II}, Proc. Roy. Soc. A 260 (1958), pp. 400--413.
\bibitem[10]{Tough82} J. T. Tough, {\em Superfluid turbulence}, in {\it Prog. Low Temp. Phys.}, ed. C. J. Gorter, North-Holland, Amsterdam, 1982,  Vol. VIII, pp. 133--220.
\bibitem[11]{Schwarz85} K. W. Schwarz, {\em Three-dimensional vortex dynamics in superfluid 4He: Line-line and line-boundary interactions}, Phys. Rev. B 31 (1985), pp. 5782--5803.
\bibitem[12]{Schwarz88} K. W. Schwarz, {\em Three-dimensional vortex dynamics in superfluid 4He: Homogenous superfluid turbulence}, Phys. Rev. B 38 (1988), pp. 2398--2417.
\bibitem[13]{Vinen02} W. F. Vinen and J. J. Niemela, {\em Quantum turbulence}, J.  Low  Temp. Phys. 128 (2002), pp. 167--231.
\bibitem[14]{Vinen06} W. F. Vinen, {\em An introduction to quantum turbulence}, J.  Low  Temp. Phys. 145 (2006), pp. 7--24.
\bibitem[15] {PLTP}  {\it Prog. Low Temp. Phys.}, eds. W. P. Halperin and M. Tsubota, Elsevier, Amsterdam, 2008, Vol. 16.
\bibitem[16]{Tsubota08} M. Tsubota, {\em Quantum turbulence}, J. Phys. Soc. Jpn. 77 (2008), 111006.
\bibitem[17]{Pethick02} C. J. Pethick and H. Smith,  {\em Bose-Einstein Condensation in Dilute Gases}, Cambridge University Press, Cambridge, 2002.
\bibitem[18]{KasamatsuPLTP} K. Kasamatsu and M. Tsubota, {\em Quantized vortices in atomic Bose-Einstein condensates}, in {\it Prog. Low Temp. Phys.}, eds. W. P. Halperin and M. Tsubota, Elsevier, Amsterdam, 2008, Vol. 16, pp. 351--403.
\bibitem[19]{Abo01} J. R. Abo-Shaeer, C. Raman, J. M. Vogels, and W. Ketterle, {\em Observation of vortex lattices in Bose-Einstein condensates}, Science 292 (2001), pp. 476--479.
\bibitem[20]{Madison00} K. W. Madison, F. Chevy, W. Wlhlleben, and J. Dalibard, {\em Vortex formation in a stirred Bose-Einstein condensate}, Phys. Rev. Lett. 84 (2000), pp. 806--809.
\bibitem[21]{Madison01} K. W. Madison, F. Chevy, W. Wlhlleben, and J. Dalibard, {\em Statonary states of  a rotating Bose-Einstein condensate: Routes to vortex nucleation}, Phys. Rev. Lett. 86 (2001), pp. 4443--4446.
\bibitem[22] {Williams74} G. A. Williams and R. E. Packard, {\em Photographs of quantized vortex lines in rotating He II}, Phys. Rev. Lett. 33 (1974), pp. 280--283.
\bibitem[23] {Yarmchuck82}E. J. Yarmchuck and R. E. Packard, {\em Photographic studies of quantized vortex lines}, J. Low Temp. Phys. 46 (1982), pp. 479--515.
\bibitem[24] {Matthews99}M. R. Matthews, B. P. Anderson, P. C. Haljan, D. S. Hall, C. E. Wieman, and E. A. Cornell, {\em Vortices in a Bose-Einstein condensate}, Phys. Rev. Lett. 83 (1999), pp. 2498--2501.
\bibitem[25]{Tsubota02}
M. Tsubota, K. Kasamatsu, and M. Ueda, {\em Vortex lattice formation in a rotating Bose-Einstein condensate}, Phys. Rev. A65 (2002), 023603.
\bibitem[26]{Kasamatsu03}
K. Kasamatsu, M. Tsubota, and M. Ueda, {\em Nonlinear dynamics of vortex lattice formation in a rotating Bose-Einstein condensate}, Phys. Rev. A67 (2003), 033610.
\bibitem[27]{Kasamatsu05a} K. Kasamatsu, M. Machida, N. Sasa and M. Tsubota, {\em Three-dimensional dynamics of vortex-lattice formation in Bose-Einstein condensate} Phys. Rev. A71 (2005), 063616.
\bibitem [28]{Frisch} U. Frisch,  {\em Turbulence},  Cambridge University Press, Cambridge, 1995.
\bibitem[29]{Kolmogorov41a} A. N.  Kolmogorov, {\em The local structure of turbulence in incompressible viscous fluid for very large Reynolds number}, Dokl. Akad. Nauk SSSR  30 (1941), pp. 299-303 [reprinted in Proc. Roy. Soc. A 434 (1991),  pp. 9-13].
\bibitem[30]{Kolmogorov41b} A. N.  Kolmogorov, {\em On degeneration (decay) of isotropic turbulence in an incompressible viscous liquid}, Dokl. Akad. Nauk SSSR  31 (1941),  pp. 538-540 [reprinted in Proc. Roy. Soc. A 434 (1991), pp. 15-17].
\bibitem[31]{Tsubota00} M. Tsubota, T. Araki, and  S. K. Nemirovskii, {\em Dynamics of vortex tangle without mutual friction in superfluid $^4$He}, Phys. Rev. B 62 (2000), pp. 11751-11762.
\bibitem[32]{Koplik93} J. Koplik and H. Levine, {\em Vortex reconnection in superfluid helium}, Phys. Rev. Lett. 71 (1993), pp. 1375-1378.
\bibitem[33]{Leadbeater01} M.  Leadbeater, T. Winiecki, D. C. Samuels, C. F. Barenghi, and C. S. Adams, {\em Sound emission due to superfluid vortex reconnections}, Phys. Rev. Lett. 86 (2001), pp. 1410-1413.
\bibitem[34]{Ogawa02} S. Ogawa, M. Tsubota, and Y. Hattori, {\em Study of reconnection and acoustic emission of quantized vortices in superfluid by the numerical analysis of the Gross-Pitaevskii equation}, J. Phys. Soc. Jpn. 71 (2002), pp. 813-821.
\bibitem[35]{Maurer98} J. Maurer and P. Tabeling, {\em Local investigation of superfluid turbulence}, Europhys. Lett. 43 (1998), pp. 29-34.
\bibitem[36]{Stalp99} S. R. Stalp, L. Skrbek, and R. J.  Donnelly, {\em Decay of grid turbulence in a finite channel}, Phys. Rev. Lett. 82 (1999), pp. 4831-4834.
\bibitem[37]{Skrbek00a} L. Skrbek, J. J. Niemela, and R. J. Donnelly, {\em Four regimes of decaying grid turbulence in a finite channel}, Phys. Rev. Lett. 85 (2000), pp. 2973-2976.
\bibitem[38]{Skrbek00b} L. Skrbek and S. R. Stalp, {\em On the decay of homogeneous isotropic turbulence}, Phys. Fluids 12 (1997), pp. 1997-2019.
\bibitem[39]{Stalp02} S. R. Stalp, J. J. Niemela, W. F. Vinen, and R. J. Donnelly, {\em Dissipation of grid turbulence in helium II}, Phys. Fluids 14 (2002), pp. 1377-1379. 
\bibitem[40]{Vinen00} W. F. Vinen, {\em Classical character of turbulence in a quantum liquid}, Phys. Rev. B 61 (2000), pp. 1410-1420.
\bibitem[41]{Nore97a} C. Nore, M. Abid, and M. E. Brachet, {\em Kolmogorov turbulence in low-temperature superflows}, Phys. Rev. Lett. 78 (1997), pp. 3296-3299.
\bibitem[42]{Nore97b} C. Nore, M. Abid, and M. E. Brachet, {\em Decaying Kolmogorov turbulence in a model of superflow}, Phys. Fluids 9 (1997), pp. 2644-2669.
\bibitem[43]{Araki02}  T. Araki, M. Tsubota and S. K. Nemirovskii, {\em Energy spectrum of superfluid turbulence with no normal-fluid component}, Phys. Rev. Lett. 89 (2002), 145301.
\bibitem[44]{Kobayashi05a} M. Kobayashi and M. Tsubota, {\em Kolmogorov spectrum of superfluid turbulence: Numerical analysis of the Gross-Pitaevskii equation with a small-scale dissipation}, Phys. Rev. Lett. 94 (2005),  065302.
\bibitem[45]{Kobayashi05b} M. Kobayashi and M. Tsubota, {\em Kolmogorov spectrum of quantum turbulence}, J. Phys. Soc. Jpn. 74 (2005), pp. 3248-3258.
\bibitem[46]{Kobayashi06a} M. Kobayashi and M. Tsubota, {\em Thermal dissipation in quantum turbulence}, Phys. Rev. Lett. 97 (2006), 145301.
\bibitem[47]{Thomson80} J. Thomson, {\em Vibrations of a columnar vortex}, Phil Mag. 10 (1880), pp. 155-168.
\bibitem[48]{Hall58} H. E. Hall, {\em An experimental and theoretical study of torsional oscillations in uniformly rotating liquid helium II}, Proc. Roy. Soc. London A 245 (1958),  pp. 546-561.
\bibitem[49]{Hall60} H. E. Hall, {\em The rotation of helium II}, Phil. Mag. Suppl. 9 (1960), pp. 89-146.
\bibitem[50]{Vinen01} W. F. Vinen, {\em Decay of superfluid turbulence at a very low temperature: The radiation of sound from a Kelvin wave on a quantized vortex}, Phys.  Rev. B  64 (2001), 134520.
\bibitem[51]{Samuels90} D. C. Samuels and R. J. Donnelly, {\em Sideband instability and recurrence of Kelvin waves on vortex cores}, Phys. Rev. Lett. 64(1990), pp. 1385-1388.
\bibitem[52]{Svistunov95} B. V. Svistunov, {\em Superfluid turbulence in the low-temperature limit,} Phys. Rev. B 52 (1995), pp. 3647-3653.
\bibitem[53]{Kivotides01} D. Kivotides, J. C. Vassilicos, D. C. Samuels, and C. F. Barenghi, {\em Kelvin waves cascade in superfluid turbulence}, Phys. Rev. Lett. 86 (2001), pp. 3080-3083.
\bibitem[54]{Vinen03} W. F. Vinen, M. Tsubota, and M. Mitani, {\em Kelvin-wave cascade on a vortex in superfluid $^4$He at a very low temperature}, Phys. Rev. Lett. 91 (2003), 135301.
\bibitem[55]{Kozik04} E. Kozik, and B. V. Svistunov, {\em Kelvin-wave cascade and decay of superfluid turbulence}, Phys. Rev. Lett. 92 (2004),  035301.
\bibitem[56]{Kozik05a} E. Kozik and B. V. Svistunov, {\em Scale-separation scheme for simulating superfluid turbulence: Kelvin-wave cascade}, Phys. Rev. Lett. 94 (2005), 025301.
\bibitem[57]{Kozik05b} E. Kozik and B. V. Svistunov, {\it Vortex-phonon interaction} Phys. Rev. B 72 (2005),  172505.
\bibitem[58]{L'vov07} V. S. L'vov, S. V.  Nazarenko, and O. Rudenko, {\em Bottleneck crossover between classical and quantum superfluid turbulence}, Phys. Rev. B 76 (2007), 024520.
\bibitem[59]{Kozik08a} E. Kozik and B. V. Svistunov, {\em Kolmogorov and Kelvin-wave cascades of superfluid turbulence at T=0: What lies between}, Phys. Rev. B 77 (2008), 060502.
\bibitem[60]{Kozik08b} E. Kozik and B. V. Svistunov, {\em Scanning superfluid-turbulence cascade by its low-temperature cutoff}, Phys. Rev. Lett. 100 (2008), 195302.
\bibitem[61]{VinenPLTP} L. Skrbek and W. F. Vinen, {\em The use of vibrating structures in the study of quantum turbulence}, in {\em Prog. Low Temp. Phys.}, eds. W. P. Halperin and M. Tsubota, Elsevier, Amsterdam, 2008,  Vol. 16, pp. 195-246.
\bibitem[62]{LandauFluid} L. D. Landau and E. M. Lifshitz, {\em Fluid mechanics}, 2nd ed., Pergamon Press, Oxford, 1987.
\bibitem[63]{DonnellyPC} R. J. Donnelly, private communication.
\bibitem[64]{Jager95}  J. J\"ager, B. Schuderer, and W. Schoepe, {\em Turbulent and laminar drag of superfluid helium on an oscillating microsphere}, Phys. Rev. Lett. 74 (1995), pp.  566-569.
\bibitem[65]{Awschalom84} D. D. Awschalom and K. W. Schwarz, {\em Observation of a remanent vorte-line density in superfluid helium}, Phys. Rev.Lett. 52 (1984), pp. 49-52.
\bibitem[66]{Hanninen07} R. H\"anninen,  M. Tsubota, and W. F. Vinen, {\em Generation of turbulence by oscillating structures in superfluid helium at very low temperatures}, Phys. Rev. B 75 (2007), 064502.
\bibitem[67]{SciverPLTP} S. W. Van Sciver and C. F. Barenghi, {\em Visualization of quantum turbulence}, in {\em Prog. Low Temp. Phys.}, eds. W. P. Halperin and M. Tsubota, Elsevier, Amsterdam, 2008,  Vol. 16, pp. 247-303.
\bibitem[68]{Zhang05} T. Zhang and  S. W. Van Sciver, {\em Large-scale turbulent flow around a cylinder in counterflow superfluid 4He (He(II))},  Nature Phys. 1 (2005), pp. 36-38.
\bibitem[69]{Poole05} D. R. Poole, C. F. Barenghi, Y. A. Sergeev, and W. F. Vinen, {\em Motion of tracer particles in He II}, Phys. Rev. B 71 (2005),  064514.
\bibitem[70]{Kivotides07} D. Kivotides, C. F. Barenghi, and Y. A. Sergeev, {\em Collision of a tracer particle and a quantized vortex in superfluid helium: Self-consistent calculations}, Phys. Rev. B 75 (2007), 212502.
\bibitem[71]{Kivotides08} D. Kivotides, C. F. Barenghi, and Y. A. Sergeev, {\em Interactions between particles and quantized vortices in superfluid helium}, Phys. Rev. B 77 (2008), 014527.
\bibitem[72]{Bewley06} G. P. Bewley, D. P. Lathrop, and K. R. Sreenivasan, {\em Visualization of quantized vortices}, Nature  441 (2006), 588.
\bibitem[73]{Paoletti08} M. S. Paelotti, R.B. Firrito, K. R. Sreenivasan, and D. P. Lathrop. {\em Visualization of superfluid helium flow}, J. Phys. Soc. Jpn. 77 (2008), 111007.
 \bibitem[74]{Kobayashi07} M. Kobayashi and M. Tsubota, {\em Quantum turbulence in a trapped Bose-Einstein condensate}, Phys. Rev. A76 (2007), 045603.
\bibitem[75]{Goto07} S. Goto, N. Ishii, S. Kida, and M. Nishioka, {\em Turbulence generator using a precessing sphere}, Phys. Fluids 19 (2007), 061705.
\bibitem[76]{Kobayashi08} M. Kobayashi and M. Tsubota, {\em Quantum turbulence in a trapped Bose-Einstein condensate under combined rotations around three axes}, J. Low Temp. Phys. 150 (2008), pp. 587-592.
\bibitem[77]{Stenger99} J. Stenger, S. Inouye, A. P. Chikkatur, D. M. Stamper-Kurn, D. E. Pritchard, and W. Ketterle, {\em Bragg spectroscopy of a Bose-Einstein condensate}, Phys. Rev. Lett. 82 (1999), pp. 4569-4573.
\bibitem[78]{Kasamatsu05b} K. Kasamatsu, M. Tsubota, and M. Ueda, {\em Vortices in multicomponent Bose-Einstein condensates}, Int. J. Mod. Phys. B 19 (2005), pp. 1835-1904.


 
\end{thebibliography}
\end{document}